# Dye-doped dual-frequency nematic cells as fast-switching polarization-independent shutters

BING-XIANG LI,[1,2] RUI-LIN XIAO,[1,2] SATHYANARAYANA PALADUGU,[1] SERGIJ V. SHIYANOVSKII,[1] AND OLEG D. LAVRENTOVICH[1, 2, 3, *]

[1]*Advanced Materials and Liquid Crystal Institute, Kent State University, Kent, Ohio, 44242, USA*
[2]*Chemical Physics Interdisciplinary Program, Kent State University, Kent, Ohio, 44242, USA*
[3]*Department of Physics, Kent State University, Kent, OH, 44242, USA*
*\*olavrent@kent.edu*

**Abstract:** We present polarization-independent optical shutters with a sub-millisecond switching time. The approach utilizes dual-frequency nematics doped with a dichroic dye. Two nematic cells with orthogonal alignment are driven simultaneously by a low-frequency or high-frequency electric field to switch the shutter either into a transparent or a light-absorbing state. The switching speed is accelerated via special short pulses of high amplitude voltage. The approach can be used in a variety of electro-optical devices.

## 1. Introduction

Anisotropic dielectric and optical properties of nematic liquid crystals (NLCs) enable a large number of electro-optical applications [1]. The dielectric anisotropy ($\Delta\varepsilon$) of an NLC determines the effect of the applied electric field $\mathbf{E}$ on the orientation of the NLC. The latter is characterized by a director $\hat{\mathbf{n}}$, which is also an optical axis of the NLC. The director aligns parallel or perpendicular to the applied electric field, depending whether $\Delta\varepsilon$ is positive or negative, respectively. The main issue of the nematic electro-optic effects is a slow relaxation time $\tau_{off} = \gamma d^2 / \pi^2 K$ when the field is switched off, typically on the order of milliseconds; here $\gamma$ is the rotational viscosity, $K$ is the elastic constant, and $d$ is the cell thickness [1]. One of the effective approaches to accelerate the process is to use the so-called dual-frequency nematic liquid crystal (DFLC) in which $\Delta\varepsilon > 0$ at field frequencies below some cross-over frequency $f_c$ and $\Delta\varepsilon < 0$ above $f_c$ [2-6]. To achieve a sub-millisecond switching time, Golovin et al [2] and Yin et al [3] proposed to use a DFLC in a high-pretilt cell. In absence of the electric field, the director makes an angle $\theta$ about 45 degrees with the normal to the substrates. When the electric field is applied, this large tilt ensures a substantial realigning torque proportional to $\sin\theta\cos\theta$ and thus a faster response time as compared to cells with either planar, $\theta = \pi/2$, or homeotropic, $\theta = 0$, alignment.

In this work, we advance the previous dual-frequency approach to construct polarization-independent sub-millisecond shutters. Shuttering is caused by the guest-host effect [7] in the DFLC cells doped with dichroic dye molecules oriented along $\hat{\mathbf{n}}$. To achieve a polarization-independent switching, we use a sandwich-type structure comprised of two identical almost homeotropic cells, in which the small in-plane director projections are orthogonal to each other, Fig. 1. In the field-off state, absorption by the dichroic dye is weak, since $\hat{\mathbf{n}}$ in both cells is nearly perpendicular to any light polarization. To create a dark state, a high-frequency voltage is applied so that $\Delta\varepsilon < 0$ and the NLC and the dye molecules realign mostly parallel to the bounding plates along the mutually orthogonal directions in the two cells. In this state, the two-cells system absorbs light strongly for all polarizations of light. To accelerate the switching from the transparent state to the light-absorbing state, a special short pulse of high amplitude and high frequency is applied prior to the high-frequency holding voltage. For a quick reverse switching, another special short pulse is used, this time of a direct current (DC) type. We achieve contrast ratio 10:1, with the transmittance changing from 48 % $\pm$ 1% in the transparent state to 5 % $\pm$ 1% in the dark state with the switching time in both directions being less than 0.3 ms.

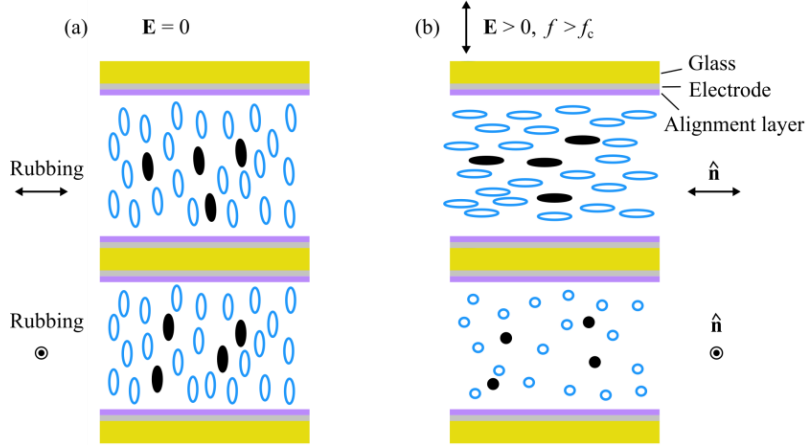

Fig. 1. Electro-optic shutter comprised of a pair of cells for polarization-independent light absorption. (a) Field-free state; the nearly homeotropic alignment of the dual frequency nematic doped with a dichroic dye makes the cells transparent to normally impinging light. The rubbing direction of two cells is perpendicular to each other. (b) Light-absorbing state is formed when a high-frequency electric field ($f > f_c$) is applied and realigns the director and dye molecules parallel to the bounding plates. The state is light-absorbing for all polarizations of light. The open and closed ellipsoids represent the nematic and dye molecules, respectively.

## 2. Experimental materials

We used homeotropic (alignment layer of polyimide SE1211, Merck) and planar (alignment layer of polyimide PI2555, HD MicroSystems) cells comprised of two glass plates with transparent indium tin oxide (ITO) electrodes of an active area $5 \times 5$ mm$^2$. The thickness $d$ of the cells was fixed by glass spheres of diameter 2-6 µm. The temperature was controlled using a Linkam LTS350 hot stage.

The switching speed of the shutter is determined by the properties of DFLCs such as rotational viscosity $\gamma$ and $\Delta\varepsilon$. We explored three DFLCs, namely, DP002-016 and DP002-026 (Jiangsu Hecheng Display Technology), and MLC-2048 (EM Industries). MLC-2048 was ruled out since its viscosity, $\gamma = 200$ mPa·s, is four times larger than that of DP002-016 ($\gamma = 48$ mPa·s) and DP002-026 ($\gamma = 51$ mPa·s) at 25°C. DP002-026 shows a crossover frequency that is significantly lower than that of DP002-016, Fig. 2(a), which implies that the cells can be driven by a lower-frequency voltages. We thus selected DP002-026 as the host material for shutters. Its dielectric permittivities, $\varepsilon_\parallel$ and $\varepsilon_\perp$, measured by using an LCR meter 4284A (Hewlett-Packard), in homeotropic and planar cells, respectively, are shown in Fig. 2(b).

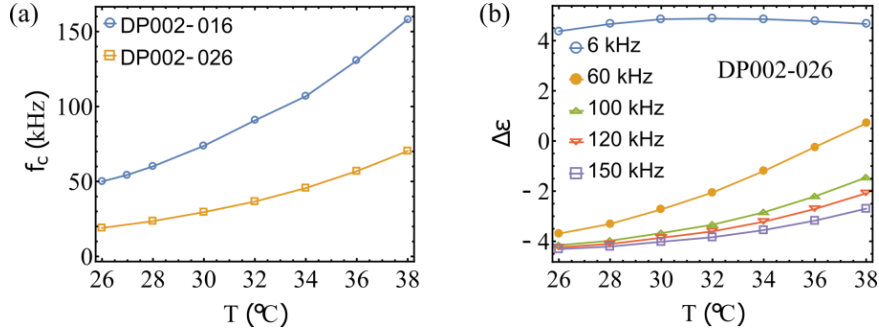

Fig. 2. Temperature dependence of (a) the crossover frequency of two studied materials, (b) dielectric anisotropy of DP002-026 at various frequencies.

The contrast ratio of the shutter is determined by the order parameter $S$ of the dye. We characterized three dichroic dyes, Sudan III (Sigma-Aldrich), AB4 (Nematel GmbH), and G-472 (Mitsui Fine Chemicals) added to DP002-026 in planar cells of 4 µm thickness at the concentration $c$=2 wt%. Each cell was probed with a linearly polarized He-Ne laser (543 nm for Sudan III, 632 nm for AB4 and G-472) at normal incidence. The transmittances of the dye-doped DFLCs are $T_\parallel = T_{cell} e^{-cd\alpha_{iso}(1+2S)}$ and $T_\perp = T_{cell} e^{-cd\alpha_{iso}(1-S)}$ for the light polarization parallel and perpendicular to the director, respectively; here $\alpha_{iso}$ is the absorption in the isotropic phase, and $T_{cell} \approx 0.9$ is the transmittance of a single cell filled with a dye-free DFLC. The order parameter is calculated from the transmission measurements, $S = \left[1 - 3\ln(T_\perp / T_{cell}) / \ln(T_\perp / T_\parallel)\right]^{-1}$, Table 1.

Table 1. Order parameter of three dichroic dyes in DP002-026.

| Material | $T_\parallel$ | $T_\perp$ | $S$ | Dye color |
| --- | --- | --- | --- | --- |
| Sudan III | 0.006 | 0.354 | 0.58 | Red |
| AB4 | 0.045 | 0.600 | 0.68 | Black |
| G-472 | 0.021 | 0.651 | 0.78 | Black |

The electro-optic cells comprising the proposed shutter shown in Fig.1 are prepared with the homeotropic alignment layer SE1211 rubbed unidirectionally in order to provide a small directional tilt of $\hat{n}$ when the electric field is absent. Each cell is assembled from pairs of plates rubbed in an antiparallel fashion. The tilt angle $\theta$ is 3.5 degrees, as determined by the crystal rotation method [8]. To achieve polarization independence for the shutter, two rubbed homeotropic cells of the same thickness are stacked such that the rubbing directions of cells were perpendicular to each other. In the field-free transparent state, the shutter transmittance is $T_{transparent} = T_\perp^2$, or $T_{transparent} = T_{cell}^2 \left(T_\parallel / T_\perp\right)^{2(1-S)/3S}$. The transmittance can be improved to $T_{transparent} = T_{cell} \left(T_\parallel / T_\perp\right)^{2(1-S)/3S}$ by using a matching fluid between the two cells. The shutter switches to an absorbing state with a low transmittance $T_{absorbing} = T_\parallel T_\perp$ when one applies a high-frequency voltage. In order to achieve $T_{transparent} = 0.55$ and the contrast ratio $T_\perp / T_\parallel = 10:1$, the order parameter should be $S \approx 0.76$ or higher, Fig. 3. Because of this, G-472 was chosen as a dye dopant, see Table 1.

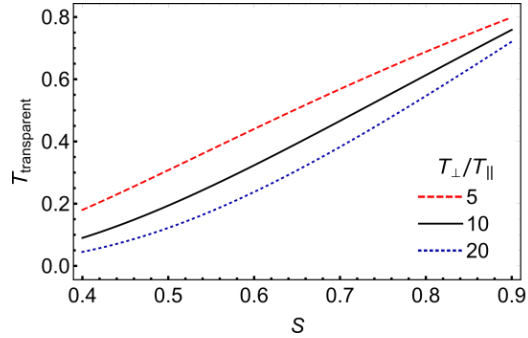

Fig. 3. Dependence of the required transmission $T_{transparent}$ at the transparent state on the minimum order parameter $S$ and contrast ratio $T_\perp / T_\parallel$.

## 3. Electro-optic performance of the shutter

To test the electro-optic performance, the cells were probed with a normally incident He-Ne laser beam (632 nm). The alternating current (AC) voltage was generated by a waveform generator (Stanford Research Systems, Model DS345) and an amplifier (Krohn-hite Corporation, Model 7602). We first characterize the performance of a single cell Fig. 4, and then the double-cell shutter, Fig. 5, filled with a mixture of DP002-026 and 2wt% of G-472.

In order to speed up the electro-optic switching, we designed a specific voltage waveform, Figs. 4 and 5. The response time for the field-induced reorientation of the nematic director is proportional to $1/E^2$ [1]. Thus, we introduced two short special pulses (SSPs) with a high voltage amplitude, Figs. 4(a), 4(b) and 5(a). The first SSP is applied to speed-up the formation of the dark planar state. This SSP is comprised of a DC pulse of duration 0.1 ms, followed by the AC pulse of frequency 60 kHz, Figs. 4(a) and 5(a). The dark state is held steady by a voltage of 7 V at 60 kHz. To switch from the dark to the transparent homeotropic state, the second SSP of a DC type was applied for 1.0 ms, after which the voltage was switched off, Figs. 4(b) and 5(a). Figures 4(c) and 4(d) show that the transmission changes are very fast, within at least ~ 0.2 ms, if one takes into account 0.1 ms delay caused by 0.1 ms DC pulse in the first SSP.

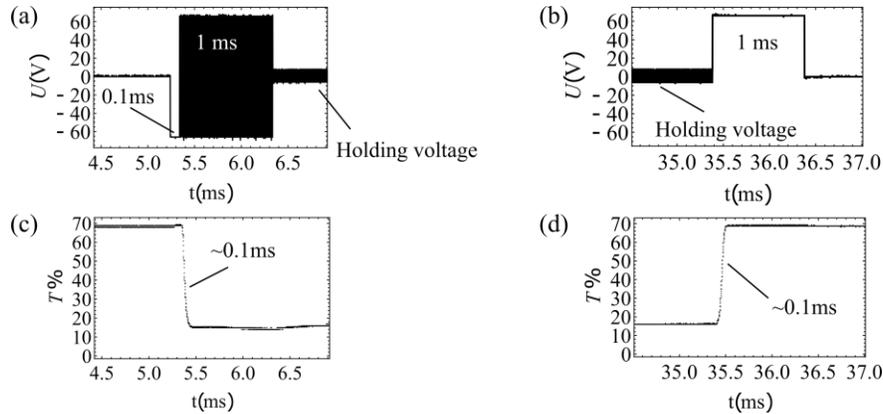

Fig. 4. Voltage waveform for transition (a) from the transparent to the dark state and (b) from the dark state to the transparent state and (c, d) the corresponding fast transmission changes measured for a single cell, $d = 5\,\mu m$, filled with DP002-026 and G-472 mixture. The polarization of incident light is along the rubbing direction of the cell. The holding voltage at 60 kHz is 7 V.

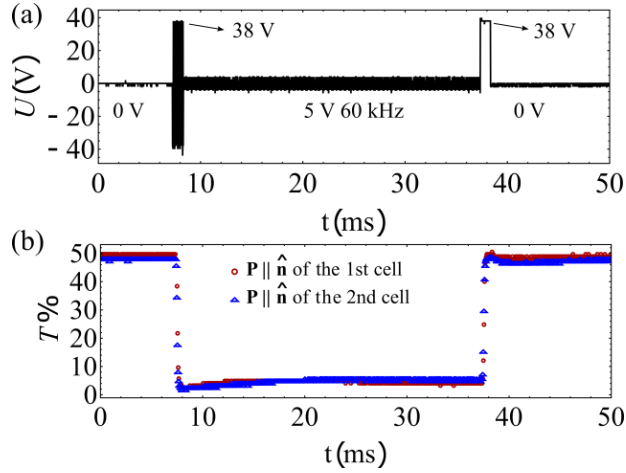

Fig. 5. (a) Voltage waveform and (b) the corresponding switching of transparency by the two-cell shutter filled with the mixture of DP002-026 and 2 wt% of G-472. The holding voltage for the dark state is 5 V at 60 kHz.

Figure 5 shows the overall performance of the shutter constructed from two cells, as in Fig. 1, each of a thickness 4.5 µm. The cells are assembled with the rubbing directions being perpendicular to each other. The gap between the cells is filled with the immersion oil (Olympus) of a refractive index 1.52. The waveform is similar to the one in Fig. 4, but with the two SSPs and the holding voltages being of a smaller amplitude, because of the smaller thickness of the cells used. The transmittance is switched from 48 % $\pm$ 1% in the transparent state to 5% $\pm$ 1% in the light-absorbing state for any polarization of incident light. The switching times are less than 0.3 ms. Depositing antireflective coatings at the external surfaces of the cells would increase the transmittance by approximately 10%, preserving the same contrast.

The contrast ratio of the device, $T_\perp / T_\parallel = \exp(3cSd\alpha_{iso})$ can be enhanced by increasing the dye concentration $c$. Of course, in some cases an increase of $c$ might lead to a decrease of $S$, but, as shown, for example, in Ref. [9], the product $cS$ generally increases with $c$. The increase of $c$ also implies a decrease of $T_{transparent}$. If the latter is not a critical issue, and can be reduced, say, by a factor of 3, then the contrast ratio would reach 1000:1. The contrast ratio can also be enhanced significantly by increasing the thickness of the cells. In this work, we intentionally selected $c$ =2wt% of G-472 in the DP002-026 in order to achieve the transmittance $T_{Transparent}$~50% of the shutter.

Another important parameter of fast switching shutters is their power consumption. To explore the issue, we consider the NLC cell as a series of a resistor $R$ (10 Ω/□) and a dielectric capacitor $C$. We determined the impedance in a cell of thickness ~4 µm and the active area $2.5 \times 2.5$ cm$^2$ filled with DP002-026 and G-472. The results show that the magnitude and the phase of the impedance of the mixture at 60 kHz are $Z \approx 450$ Ω and $\phi \approx 70°$, respectively. One can estimate the power for the holding voltage and the special short pulse at 60 kHz, i.e. $P_h = U_h^2 \cos\phi / Z \approx 0.02$ W and $P_S = U_S^2 \cos\phi / Z \approx 1.2$ W, where the typical holding voltage is $U_h \approx 5$ V and the voltage of the special pulse is $U_S = 40$ V. The energy consumed by special DC and AC pulses (that are very short in duration, ~0.3 ms) used to accelerate the switching, is less than 1 mJ. The power consumption of the display of a smart phone is typically ~ 0.5 W [10]. Thus, the power consumption should not be a problem for the practical application.

One of the potentially detrimental features of DFLC-based devices is dielectric heating. According to Refs. [11, 12], it increases the temperature of DFLC by $\Delta T \approx \pi f \varepsilon_0 \varepsilon'' U^2 / hd$, where $\varepsilon_0 = 8.85 \times 10^{-12}$ F/m is the vacuum permittivity, $\varepsilon''$ is the imaginary part of the permittivity of DFLCs, $U$ is the applied voltage and $h$ is the heat transfer coefficient of the surrounding medium. We measured the real and imaginary parts of permittivity of DP002-026 as $\varepsilon' \approx 8.8$ and $\varepsilon'' \approx 0.03$ at 60 kHz in the planar state at 28°C. With the heat transfer coefficient of the still air being $h = 16$ Wm$^{-2}$K$^{-1}$ [11, 12], one can estimate that the temperature of DP002-026 in a cell with $d = 4.5$ μm will increase by 0.02°C and 1.1°C for the applied voltages 5 V and 40 V, respectively, at 60 kHz. Such a small heating will not significantly change the dielectric anisotropy of the material and thus would not be detrimental to the shutter performance.

## 4. Conclusions

We demonstrate polarization-independent fast electro-optical switching of a two-cell shutter based on a dual-frequency nematic doped with a dichroic dye. The demonstrated electro-optic effect may be used in switchers, eyewear, and color filters. The response time is less than 0.3 ms for both transparent-dark and dark-transparent transitions. The power consumption and dielectric heating effects are not significant. The proposed optical scheme of the shutter can be implemented in other geometries. For example, the two homeotropic cells can be replaced by two planar cells with the director fields that are mutually perpendicular. In this geometry, the field-free state is dark. A low frequency field would realign the director in both cell homeotropically and the device would become transparent.

**Funding**

The work was partially supported by Valeo, Inc. (France).

**Acknowledgment**

The authors acknowledge useful discussions with Dr. Kedar Sathaye.

## References

1. D.-K. Yang and S.-T. Wu, *Fundamentals of liquid crystal devices,* II ed. (John Wiley & Sons, 2014).
2. A. B. Golovin, S. V. Shiyanovskii, and O. D. Lavrentovich, "Fast switching dual-frequency liquid crystal optical retarder, driven by an amplitude and frequency modulated voltage," Appl. Phys. Lett. **83**, 3864-3866 (2003).
3. Y. Yin, M. Gu, A. Golovin, S. Shiyanovskii, and O. Lavrentovich, "Fast switching optical modulator based on dual frequency nematic cell," Mol. Cryst. Liq. Cryst. **421**, 133-144 (2004).
4. M. Mrukiewicz, P. Perkowski, W. Piecek, R. Mazur, O. Chojnowska, and K. Garbat, "Two-step switching in dual-frequency nematic liquid crystal mixtures," J. Appl. Phys. **118**, 173104 (2015).
5. X.-W. Lin, W. Hu, X.-K. Hu, X. Liang, Y. Chen, H.-Q. Cui, G. Zhu, J.-N. Li, V. Chigrinov, and Y.-Q. Lu, "Fast response dual-frequency liquid crystal switch with photo-patterned alignments," Opt. Lett. **37**, 3627-3629 (2012).
6. W. Duan, P. Chen, B.-Y. Wei, S.-J. Ge, X. Liang, W. Hu, and Y.-Q. Lu, "Fast-response and high-efficiency optical switch based on dual-frequency liquid crystal polarization grating," Opt. Mater. Express **6**, 597-602 (2016).
7. G. H. Heilmeier and L. A. Zanoni, "Gust-host interactions in nematic liquid crystals. A new electro-optic effect," Appl. Phys. Lett. **13**, 91-92 (1968).
8. K. Y. Han, T. Miyashita, and T. Uchida, "Accurate measurement of the pretilt angle in a liquid crystal cell by an improved crystal rotation method," Mol. Cryst. Liq. Cryst. **241**, 147-157 (1994).
9. A. Ranjkesh, J.-C. Choi, K.-I. Joo, H.-W. Park, M. S. Zakerhamidi, and H.-R. Kim, "Linear dichroism and order parameters of nematics doped with azo dyes," Mol. Cryst. Liq. Cryst. **647**, 107-118 (2017).
10. S. Tarkoma, M. Siekkinen, E. Lagerspetz, and Y. Xiao, *Smartphone energy consumption: modeling and optimization* (Cambridge University, 2014).
11. Y. Yin, S. V. Shiyanovskii, and O. D. Lavrentovich, "Electric heating effects in nematic liquid crystals," J. Appl. Phys. **100**, 024906 (2006).
12. Y.-C. Hsiao and W. Lee, "Lower operation voltage in dual-frequency cholesteric liquid crystals based on the thermodielectric effect," Opt. Express **21**, 23927-23933 (2013).